\documentclass[10pt,letterpaper]{article}
\usepackage{opex3}
\usepackage{amstext}
\usepackage{graphicx}
\usepackage{bm}
\usepackage{amssymb}
\usepackage{amsthm}
\usepackage{float}

\begin{document}

\title{100~km secure differential phase shift quantum key distribution with low jitter up-conversion detectors}

\author{Eleni Diamanti$^1$, Hiroki Takesue$^{2,3}$, Carsten Langrock$^1$, \\M. M. Fejer$^1$ and Yoshihisa Yamamoto$^1$}

\address{$^1$Edward L. Ginzton Laboratory, Stanford University, Stanford, California 94305-4088}

\address{$^2$NTT Basic Research Laboratories, NTT Corporation, 3-1 Morinosato Wakamiya, Atsugi, 243-0198, Japan}

\address{$^3$CREST, Japan Science and Technology Agency, 4-1-8 Honcho, Kawaguchi, Saitama, 332-0012, Japan}

\email{ediam@stanford.edu}

\begin{abstract}
We present a quantum key distribution experiment in which keys that
were secure against all individual eavesdropping attacks allowed by
quantum mechanics were distributed over 100~km of optical fiber. We
implemented the differential phase shift quantum key distribution
protocol and used low timing jitter 1.55~$\mu$m single-photon
detectors based on frequency up-conversion in periodically poled
lithium niobate waveguides and silicon avalanche photodiodes. Based
on the security analysis of the protocol against general individual
attacks, we generated secure keys at a practical rate of 166~bit/s
over 100~km of fiber. The use of the low jitter detectors also
increased the sifted key generation rate to 2~Mbit/s over 10~km of
fiber.
\end{abstract}


\section{Introduction}

Since the first demonstration of a quantum key distribution (QKD)
system in 1992~\cite{bennett:jcrypto92}, there have been numerous
efforts toward the implementation of such systems~\cite{gisin:rmp02}
with the goal of making quantum cryptography practical by achieving
the longest possible communication distance and the highest possible
communication rate. The rapid progress in the field has recently led
to the implementation of fiber-based QKD systems that operated at
1~GHz clock frequency~\cite{honjo:ol04,takesue:njp05} and extended
the key distribution distance to more than
100~km~\cite{takesue:njp05,gobby:apl04,rosenberg:quantph06}.

Most of the previous experiments, however, have not been able to
guarantee the security of the generated keys against general
eavesdropping attacks. Often the average photon number per pulse is
set to the arbitrary value 0.1, which is not the result of a
security proof. In the best cases, only a limited set of potential
eavesdropping attacks is taken into account. For implementations of
the BB84 protocol~\cite{bennett:84} with a Poisson source this set
usually does not include the powerful photon number splitting
attack, rendering these systems ultimately
insecure~\cite{lutkenhaus:pra00,brassard:prl00}. Generation of keys
that were secure against general individual attacks and their
distribution over 50~km of optical fiber using the BB84 protocol
with a Poisson source and InGaAs/InP avalanche photodiodes (APDs)
with a very small dark count rate was reported in~\cite{gobby:el04}.
Furthermore, \cite{rosenberg:quantph06} and \cite{zhao:quantph}
reported implementations of the decoy state BB84
protocol~\cite{lo:prl05,wang:prl05}, which achieved secure key
distribution over 107~km and 60~km of fiber using superconducting
transition-edge sensors and InGaAs/InP APDs, respectively. All these
systems, however, featured a very small secure key generation rate
of $<$1~bit/s, which prevents their integration into practical
telecommunication networks.

In the experiment presented in~\cite{takesue:njp05}, we implemented
the differential phase shift quantum key distribution (DPS-QKD)
protocol, which uses a Poisson light
source~\cite{inoue:prl02,inoue:pra03} but is robust to photon number
splitting attacks~\cite{takesue:njp05,diamanti:pra05,inoue:pra05}.
In this implementation, the security analysis against a limited set
of eavesdropping attacks, in particular the beamsplitter,
intercept-resend and photon number splitting attacks, was taken into
account. Although it is important to demonstrate a practical QKD
system that is secure against these realistic attacks, it is also
crucial to consider more elaborate attacks that will be within
technological reach in the near future and guarantee the security of
the system against these types of attacks. The security of the
DPS-QKD protocol against all individual attacks allowed by quantum
mechanics, including photon number splitting attacks, was proven
in~\cite{waks:pra06}. Based on this security analysis, the
experiment in~\cite{takesue:njp05} did not generate secure keys over
105~km of fiber. The main limiting factor for the secure key
distribution distance in this system was the large bit error rate
caused by the broadening of the received signal induced by the large
timing jitter of the single-photon detectors employed in the system.
These detectors were based on frequency up-conversion in
periodically poled lithium niobate (PPLN) waveguides and Si
APDs~\cite{langrock:ol05}, and had a jitter of $\sim$~500 ps.
However, up-conversion detectors using Si APDs with improved timing
jitter characteristics were recently reported~\cite{thew:njp06}.

In this paper, we use the security analysis of the DPS-QKD protocol
against general individual attacks and low jitter up-conversion
detectors to implement a secure high speed and long distance quantum
key distribution system. The use of the low jitter detectors
significantly improved the signal to noise ratio, which resulted in
a smaller bit error rate. Thus, despite the tight security
requirements, we achieved the distribution of keys that were secure
against all individual attacks allowed by quantum mechanics over
100~km of optical fiber at a rate of 166~bit/s, which is two orders
of magnitude higher than previously reported values. Furthermore,
using the low jitter detectors allowed us to increase the sifted key
generation rate to 2~Mbit/s over 10~km of fiber, which is double
than the previous record~\cite{takesue:njp05}.

\section{Security of the differential phase shift quantum key distribution protocol}

\begin{figure}[htbp]
\centering\includegraphics[width=13cm]{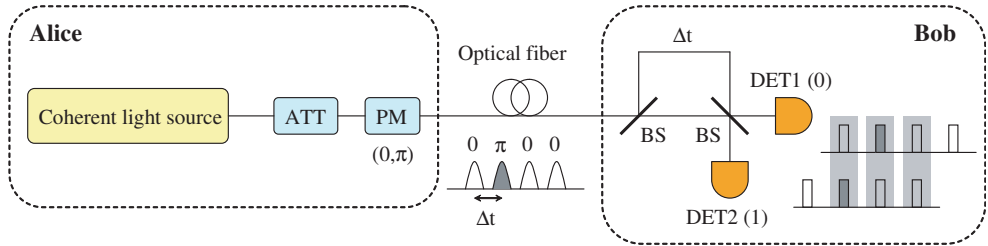}
\caption{Quantum key distribution system for the implementation of
the DPS-QKD protocol. ATT, attenuator; PM, phase modulator; BS,
beamsplitter; DET, detector.} \label{fig:dps-setup}
\end{figure}

A quantum key distribution system that implements the DPS-QKD
protocol is shown in Fig.~\ref{fig:dps-setup}. Alice generates a
train of coherent pulses, which are attenuated such that the average
photon number per pulse is less than 1, randomly phase modulated by
0 or $\pi$, and sent over an optical fiber to Bob. Each photon
coherently spreads over many pulses with a fixed phase modulation
pattern. In the receiver side, Bob divides the incoming pulses into
two paths and recombines them using 50/50 beamsplitters. The time
delay introduced by his interferometer is equal to the inverse of
the clock frequency, or else equal to the time separation between
sequential pulses. Single-photon detectors are placed at the output
ports of the second beamsplitter. After passing through Bob's
interferometer, the pulses interfere at the output beamsplitter and
the phase difference between two consecutive pulses determines which
detector records a detection event. Detector 1 in
Fig.~\ref{fig:dps-setup} records an event when the phase difference
is 0 and detector 2 records an event when the phase difference is
$\pi$. Because the average photon number per pulse is less than one,
Bob observes detection events only occasionally and at random time
instances. Bob announces publicly the time instances at which a
photon was detected, but he does not reveal which detector detected
it. From her modulation data, Alice knows which detector in Bob's
site recorded the event. Thus, by assigning bit values 0 and 1 to
detection events recorded by detector 1 and 2, respectively, they
form a secret key.

In general terms, the security of the DPS-QKD protocol stems from
the nondeterministic collapse of a wavefunction in a quantum
measurement. In particular, if the number of pulses in the coherence
time of Alice's source is $n_{p}$, then each of Alice's photons is
in a superposition of all the states that correspond to the $n_{p}$
time instances with the appropriate phase applied to each one of
them. The overall wavefunction is a product state of these
individual photon states. At Bob's site, a detection event at a
certain time instance $t_n$ reveals the phase difference between the
pulses in time instances $t_n$ and $t_{n+1}$, which corresponds to
one bit of information. However, these detection events occur
completely randomly, so an eavesdropper cannot deterministically
collapse the wavefunction in the same time instance and obtain the
same bit of information as Bob.

The security of the DPS-QKD protocol against general individual
attacks was rigorously proven in~\cite{waks:pra06}. This analysis
considered a twofold eavesdropping strategy. Eve, the eavesdropper,
measures the photon number in the $n_p$-slot wavefunction using a
quantum non-demolition (QND) measurement. Then, she sends to Bob
$n_p\mu T$ photons, where $\mu$ is the average photon number per
pulse and $T$ is the total transmission efficiency of the quantum
channel and Bob's detection setup, and she stores $n_p\mu(1-T)$
photons coherently to be measured after Alice and Bob have revealed
all classical information. This is the photon number splitting
attack in the case of the DPS-QKD protocol. In the case that Eve is
assumed to store and measure her photons individually it was shown
in~\cite{waks:pra06} that she can obtain complete information for a
fraction $2\mu(1-T)$ of the sifted key. When $T\ll 1$ and $\mu$ is
small this attack is relatively ineffective for the DPS-QKD
protocol. However, in the presence of system errors, Eve can also
apply an optimal measurement attack on a fraction of the photons
transmitted to Bob. Assuming that Eve attaches an individual probe
state to each single photon, and then measures the probes
independently after all classical information has been revealed, it
was shown that the collision probability for each bit
$p_{\text{c}_0}$ is bounded as follows~\cite{waks:pra06}:
\begin{equation}
p_{\text{c}_0}\leq 1-e^2-\frac{\left(1-6 e\right)^2}{2}
\label{eq:dps-pcind}
\end{equation}
where $e$ is the innocent system error rate.

Taking into account the results of the photon splitting and general
individual attacks analysis, the average collision probability for
the $n$-bit sifted key, which is a measure of Eve's mutual
information with Alice and Bob, is given by the expression:
\begin{equation}
p_{\text{c}}=p_{\text{c}_0}^n=\left[1-e^2-\frac{\left(1-6
e\right)^2}{2}\right]^{n\left[1-2\mu(1-T)\right]}
\label{eq:general-pc}
\end{equation}
Then, the shrinking factor applied during privacy amplification to
guarantee the security of the generated key is calculated as
follows:
\begin{equation}
\tau=-\frac{\log_2
p_{\text{c}}}{n}=-\left[1-2\mu(1-T)\right]\log_2\left[1-e^2-\frac{\left(1-6
e\right)^2}{2}\right] \label{eq:tauDPS-general}
\end{equation}
Finally, using the techniques of the generalized privacy
amplification theory the secure key generation rate after error
correction and privacy amplification is given by the
expression~\cite{lutkenhaus:pra00}:
\begin{eqnarray}
R_{\text{secure}}&=&R_{\text{sifted}}\left\{\tau+f(e)\left[e\log_2
e+(1-e)\log_2(1-e)\right]\right\} \nonumber \\
&=&R_{\text{sifted}}\{-[1-2\mu(1-T)]\log_2[1-e^2-\frac{\left(1-6
e\right)^2}{2}] \nonumber \\
 & & + f(e)[e\log_2 e+(1-e)\log_2(1-e)]\} \label{eq:1GHz-RDPSind}
\end{eqnarray}
where the sifted key generation rate $R_{\text{sifted}}$ is given by
Eq.~(\ref{eq:up-Rrawng}) below for the up-conversion single-photon
detectors and $f(e)$ characterizes the performance of the error
correction algorithm. The QKD experiments presented in
Sec.~\ref{sec:experiment} are based on the results of the security
analysis described here, and in particular
Eq.~(\ref{eq:1GHz-RDPSind}).

\section{The low jitter up-conversion detector}

In the 1.55~$\mu$m up-conversion single-photon
detector~\cite{langrock:ol05}, a single photon at 1.55~$\mu$m is
combined with a strong pump at 1.32~$\mu$m in a wavelength division
multiplexing coupler, and subsequently the two beams interact in a
PPLN waveguide, designed for sum frequency generation at these
wavelengths. This device allows for an internal conversion
efficiency exceeding 99\% of the signal to the 713~nm sum frequency
output. After a long-pass filter, a dichroic beamsplitter and a
prism that serve the purpose of eliminating the residual pump and
its second harmonic, the converted photon is detected by a Si APD.
The up-conversion detector presents more favorable characteristics
for fiber-based quantum cryptography than the commonly used
InGaAs/InP APD~\cite{diamanti:pra05}. This is mainly because Si APDs
have a low afterpulse probability, which enables free-running or
nongated Geiger mode operation. Thus, the sifted key generation rate
in the DPS-QKD system is only limited by the dead time of the Si
APD, and is written as:
\begin{equation}
R_{\text{sifted}}=\nu\mu Te^{-\nu\mu T t_d/2} \label{eq:up-Rrawng}
\end{equation}
where $t_d$ is the detector dead time, $\nu$ the system clock
frequency, and the factor $1/2$ in the exponent appears because the
average number of photons per second that reach each detector in
Bob's setup is $\nu\mu T/2$. For commercial Si APDs with a dead time
on the order of 50-80~ns, the exponential term becomes appreciable
for low fiber losses and high count rates. The nongated mode
operation, however, does not impose any severe limitation on the QKD
system clock frequency, which is only determined by the speed of the
electronic equipment and the Si APD timing jitter. In the
experiments described in this paper a clock frequency of 1~GHz was
used, while a 10~GHz system is also possible with these
detectors~\cite{takesue:oe06}.

\begin{figure}[htbp]
\centering\includegraphics[width=10cm]{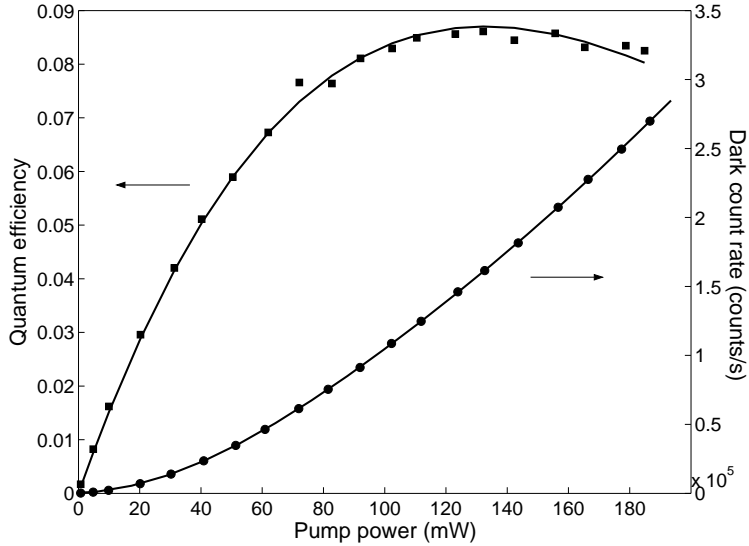}
\caption{Quantum efficiency and dark count rate of the low jitter
up-conversion detector as a function of pump power.}\label{fig:qedc}
\end{figure}

The quantum efficiency and dark count rate experimental data for the
up-conversion single-photon detectors with the low jitter Si APDs
(MPDs) that were used for the QKD experiments are shown in
Fig.~\ref{fig:qedc}. The quantum efficiency of the MPD device at the
output signal wavelength of 713~nm is $\sim 25$\%, and so the
maximum quantum efficiency of the up-conversion detector, including
the coupling, propagation, and collection setup losses, did not
exceed 9\% for 130~mW of pump power. The dark counts, on the other
hand, increase approximately quadratically with the pump power
because of parasitic nonlinear processes in the waveguide and the
input fiber~\cite{langrock:ol05}.

\begin{figure}[htbp]
\centering\includegraphics[width=10cm]{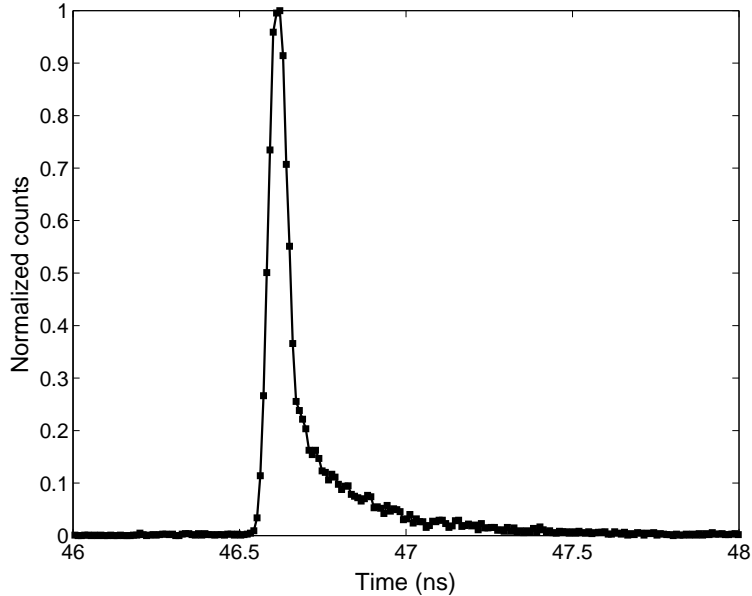}
\caption{Typical detection signal from the low jitter up-conversion
detector when 66~ps pulses are used. This curve corresponds to a
count rate of $10^5$~counts/s.}\label{fig:jitter}
\end{figure}

In order to evaluate the performance of the low jitter up-conversion
detectors for the DPS-QKD system we perform timing jitter
measurements. For these measurements, pulses with a full width at
half maximum (FWHM) of 66~ps at a repetition rate of 100~MHz are
sent to the detector and the detection signal is recorded with a
time interval analyzer. Under these conditions, a typical detection
signal from the up-conversion single-photon detector with the low
jitter Si APD is shown in Fig.~\ref{fig:jitter} for a count rate of
$10^5$~counts/s. As we observe in this figure, the FWHM is 75~ps,
which is significantly smaller than the 500~ps jitter obtained in
experiments with high jitter up-conversion detectors. Nevertheless,
the detection signal is clearly not Gaussian; there is a tail that
can potentially cause errors in the adjacent 1~ns time slot in a
DPS-QKD experiment with a clock frequency of 1~GHz.
Fig.~\ref{fig:jitter} shows, however, that 1~ns away from the peak
the tail is sufficiently small to prevent intersymbol interference.
It is clear that the improvement in timing jitter achieved with the
low jitter Si APDs is significant, and so the error rate should be
considerably lower in QKD systems employing these detectors.

\section{\label{sec:experiment}DPS-QKD experimental setup and results}

\begin{figure}[htbp]
\centering\includegraphics[width=13cm]{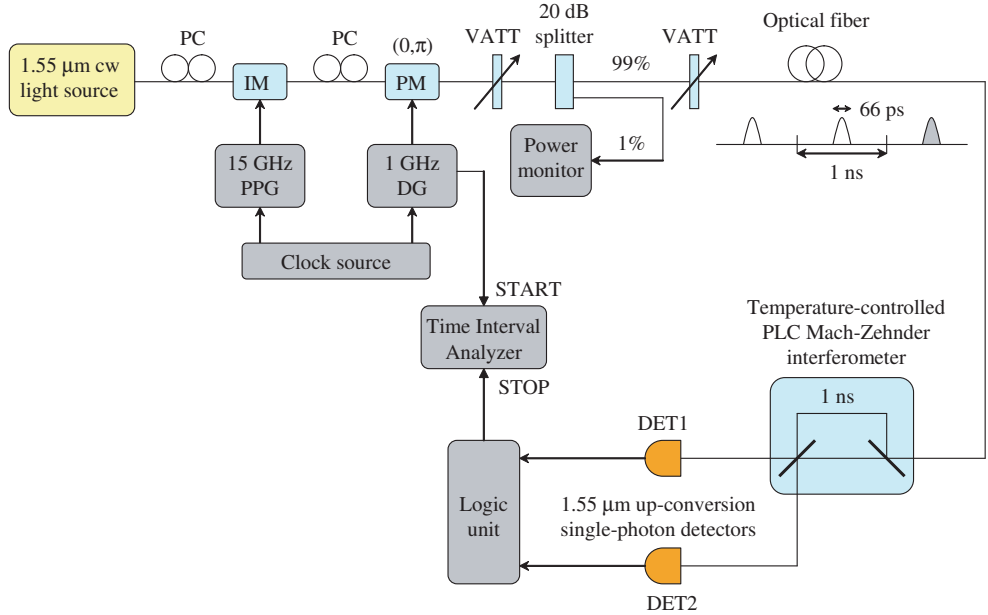}
\caption{Experimental setup for the 1 GHz DPS-QKD system. PC,
polarization controller; IM, intensity modulator; PM, phase
modulator; VATT, variable attenuator; PPG, pulse pattern generator;
DG, data generator.}\label{fig:qkd-setup}
\end{figure}

The experimental setup for the quantum key distribution experiments
that we performed at a 1~GHz clock frequency to implement the
DPS-QKD protocol with low jitter up-conversion single-photon
detectors is shown in Fig.~\ref{fig:qkd-setup}. At Alice's site, a
continuous wave light at 1.55~$\mu$m generated from an external
cavity semiconductor laser was modulated into a coherent pulse train
with a 1~GHz clock frequency using a $\text{LiNbO}_3$ intensity
modulator. The modulator was driven by a 15~GHz pulse pattern
generator, so the pulse width was 66~ps. Subsequently, following the
DPS-QKD protocol that is illustrated in Fig.~\ref{fig:dps-setup},
the phase of each pulse was modulated by 0 or $\pi$ with a
$\text{LiNbO}_3$ phase modulator. The phase modulation signal was a
1~Gbit/s pseudo-random bit sequence with a length of $2^7-1$ bits,
which was generated by a data generator. The pulses were
appropriately attenuated and sent to Bob's site through an optical
fiber, where a 1-bit delay Mach-Zehnder interferometer based on
planar lightwave circuit (PLC) technology was installed. The
insertion loss of the interferometer was 2~dB, and the extinction
ratio was greater than 20~dB. One 1.55~$\mu$m up-conversion
single-photon detector was connected to each of the output ports of
the interferometer. The events detected by the two Si APDs were
recorded using a time interval analyzer.

In order to reduce the bit error rate caused by the large dark
counts of the up-conversion detector we set the pump power at
relatively low levels, at the expense of reduced quantum efficiency
and thus reduced key generation rate as well. To further reduce the
bit error rate due to dark counts we applied a time window to the
recorded data. Because of the improved timing jitter characteristics
of the up-conversion detector, which induces a small pulse
broadening as illustrated in Fig.~\ref{fig:jitter}, the signal
counts are concentrated in small time segments while the dark counts
are randomly distributed. Therefore, we can use short measurement
time windows to reduce the effective dark counts and improve the
signal to noise ratio. This results in a significantly smaller bit
error rate.

\begin{figure}[htbp]
\centering\includegraphics[width=10cm]{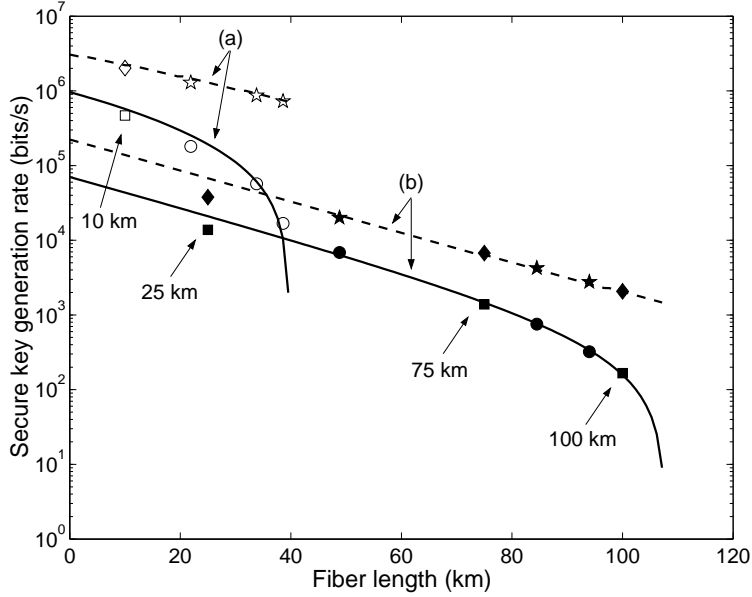} \caption{Secure
and sifted key generation rate as a function of fiber length for two
cases. (a) The dashed and solid curves are theoretical predictions
for the sifted and secure rate, respectively, when $\eta=6\%$ and
$d=1.95\times 10^{-5}$. The clear diamond and square are the
experimental fiber transmission data for the sifted and secure key
generation rate under these conditions. The clear stars and circles
are the data taken with attenuation used to simulate additional
fiber loss. (b) The dashed and solid curves are the theoretically
predicted sifted and secure rate, when $\eta=0.4\%$ and $d=3.5\times
10^{-8}$. The filled diamonds and squares are the experimental fiber
transmission data under these conditions. The filled stars and
circles are the simulated attenuation data. A baseline system error
rate of 1.5\% is assumed in all theoretical
calculations.}\label{fig:data}
\end{figure}

Before performing QKD experiments, we set the average photon number
per pulse $\mu$ at its optimal value. In particular, based on the
experimental parameters of the system, we maximized the secure key
generation rate with respect to $\mu$ using
Eq.~(\ref{eq:1GHz-RDPSind}) which corresponds to the general
individual attacks security analysis. The optimal value was 0.2.
Subsequently, we performed QKD experiments, that is we measured the
generation rate of the sifted keys that Alice and Bob exchanged, and
by directly comparing the yielded keys we also measured the bit
error rate of the transmission. For each fiber length, we measured
the sifted key generation rate and error rate five times and took
the average values. We then calculated the secure key generation
rate from Eq.~(\ref{eq:1GHz-RDPSind}) using the experimental results
for the sifted key generation rates and bit error rates. Alice and
Bob were located in the same room and we performed fiber
transmission experiments using fiber spools, while some additional
data were taken with an optical attenuator simulating fiber loss.
Fig.~\ref{fig:data} shows the theoretical curves and experimental
results for the sifted and secure key generation rate as a function
of fiber length that we obtained with the described setup and
procedure for two different experimental conditions.

We first set the detector operating condition to levels appropriate
for achieving high speed quantum key distribution over a short
communication distance. More specifically, the quantum efficiency
and dark count rate of the low jitter up-conversion detectors were
set to 6\% and 98~kHz, respectively. These values do not correspond
to the same pump power level in Fig.~\ref{fig:qedc} because the
performance of the detectors was slightly degraded when the QKD
experiments were performed compared to when the quantum efficiency
and dark count rate data were taken. We set the time window width to
200~ps, so the dark counts per time window in these experiments were
$d=1.95\times 10^{-5}$. The use of the 200~ps time window also
decreased the effective quantum efficiency by 40\%. Under these
operating conditions, we performed QKD experiments for 10~km of
optical fiber. The curves (a) of Fig.~\ref{fig:data} correspond to
the theoretical prediction for the sifted and secure key generation
rate under these experimental conditions, when $\mu$ is optimized to
maximize the secure key generation rate using the general individual
attacks security analysis. A baseline system bit error rate of 1.5\%
was assumed in these calculations. The clear square represents the
fiber transmission experimental result for the secure key generation
rate, while the sifted key generation rate at the corresponding
fiber length is represented by the clear diamond. The clear circles
and stars show the experimental results when we simulated additional
fiber loss with an optical attenuator. As we observe in
Fig.~\ref{fig:data}, the theoretical curves fit very well with the
experimental results. At the fiber length of 10~km we achieved a
sifted key generation rate of 2~Mbit/s with a bit error rate of
2.2\%, thus the secure key generation rate at this fiber length was
0.468~Mbit/s. The use of the low jitter detectors resulted in a
double sifted rate at small fiber loss compared to previous
experiments with high jitter up-conversion
detectors~\cite{takesue:njp05} because of the significantly reduced
error rate.

Subsequently, we set the quantum efficiency, dark count rate, and
time window width to 0.4\%, 350~Hz and 100~ps, respectively, to
further reduce the errors caused by dark counts and thus improve the
signal to noise ratio to achieve long distance quantum cryptography.
The use of the 100~ps time window set the dark counts per time
window to $d=3.5\times 10^{-8}$, and also reduced the effective
quantum efficiency of the detector by 54\%. Under these operating
conditions, we performed QKD experiments for 25, 75 and 100~km of
optical fiber. The curves (b) of Fig.~\ref{fig:data} correspond to
the theoretical prediction for the sifted and secure key generation
rate when the above experimental conditions are assumed. The filled
squares and diamonds represent the fiber transmission experimental
results, while the filled circles and stars correspond to data taken
using the attenuator to simulate additional fiber loss. Again, we
observe that the theoretical curves fit very well with the
experimental data. By using these operating conditions, keys that
were secure against general individual eavesdropping attacks were
distributed at a rate of 166~bits/s over 100~km of fiber. The bit
error rate for the 100~km experiment was 3.4\%, of which 1\% is
attributed to imperfect interferometry, 1.7\% to detector dark
counts, and the remaining 0.7\% to the timing jitter. This result
shows that the key distribution distance for which security against
all individual attacks allowed by quantum mechanics is guaranteed
for the DPS-QKD protocol was considerably extended because of the
improved timing jitter characteristics of the up-conversion
detectors employed in the system. These characteristics led to small
pulse broadening, which allowed the use of a short measurement time
window to substantially reduce the effective dark counts, thus
improving the signal to noise ratio and decreasing the bit error
rate.

\section{Conclusion}

We presented a practical and secure quantum key distribution system
that implemented the differential phase shift quantum key
distribution protocol with low jitter up-conversion detectors. We
showed that the improved timing jitter characteristics of the
detectors allowed us to significantly increase both the key
distribution rate and distance of the DPS-QKD system, while at the
same time guaranteeing its security against the most general
individual eavesdropping attacks allowed by quantum mechanics. With
this system we achieved a 2~Mbit/s sifted key generation rate with a
corresponding secure key generation rate of 0.468~Mbit/s over 10~km
of optical fiber, and secure key distribution over 100~km of fiber
at a rate of 166~bit/s, which is two orders of magnitude higher than
previously reported values.

The quantum cryptography system we presented achieves a sufficiently
high communication rate and a long enough communication distance to
be able to operate in a standard telecommunication network. However,
the system's capabilities can be further extended by improving the
dark count behavior of the up-conversion detectors and the timing
jitter characteristics of the Si APDs. The dark counts caused by
noise photons generated via spontaneous Raman scattering can be
reduced by using a shorter signal wavelength than pump wavelength,
while single-photon detectors with a Gaussian response and narrow
FWHM based for example on photomultiplier tubes may soon become
available. This will open the way to megahertz secure key generation
rates and very long distance secure communication.\\
\\
{\bf Acknowledgements}\\
\\
The authors would like to thank E. Waks for useful comments and G.
Kalogerakis for providing fiber spools and electronic components. H.
Takesue also thanks T. Honjo and Y. Tokura for helpful discussions
and support during his stay at Stanford University. Financial
support was provided by the SORST program of Japan Science and
Technology Agency (JST), the National Institute of Information and
Communications Technology (NICT) of Japan, and the MURI Center for
Photonic Quantum Information Systems (ARO/ARDA DAAD19-03-1-0199).

\end{document}